\documentclass[prl,twocolumn,floats,aps,epsfig,nofootinbib,amssymb]{revtex4}
\usepackage{subfigure}
\usepackage[dvips]{graphicx}
\usepackage{amsmath}
\usepackage{bm}
\usepackage{times}
\usepackage{epsfig}
\usepackage{multirow}
\newcommand{\be}{\begin{equation}}
\newcommand{\ee}{\end{equation}}
\newcommand{\bea}{\begin{eqnarray}}
\newcommand{\eea}{\end{eqnarray}}

\newcommand{\bi}{\begin{itemize}}
\newcommand{\ei}{\end{itemize}}
\newcommand{\bmt}{\begin{pmatrix}}
\newcommand{\emt}{\end{pmatrix}}
\newcommand{\bt}{\begin{tabular}}
\newcommand{\et}{\end{tabular}}

\newcommand{\benu}{\begin{enumerate}}
\newcommand{\eenu}{\end{enumerate}}

\newcommand{\ba}{\begin{array}{c}}
\newcommand{\baz}{\begin{array}{cc}}
\newcommand{\bad}{\begin{array}{ccc}}
\newcommand{\bav}{\begin{array}{cccc}}
\newcommand{\ea}{\end{array}}

\begin{document}
\title{Neutrinoless double beta decay and pseudo-Dirac neutrino mass predictions \\
through inverse seesaw mechanism} 
\author{{\bf Ram Lal Awasthi,$^\delta$ ~M. K. Parida$^\dagger$ and Sudhanwa Patra$^\dagger$}\\
\vspace*{0.5cm}
{\it \small $^\dagger$Center of Excellence in Theoretical and Mathematical Sciences, \\
Siksha 'O' Anusandhan University, Bhubaneswar-751030, India.}\\
{\it \small $^\delta$Harish-Chandra Research Institute, Chhatnag Road, Jhusi, Allahabad 211019, India.}}
\email{ ramlal@hri.res.in, sudha.astro@gmail.com, paridamk@soauniversity.ac.in}

\begin{abstract}

In the inverse seesaw extension of the standard model, supersymmetric or non-supersymmetric, 
while the light left-handed neutrinos are Majorana, the heavy right-handed neutrinos are 
pseudo-Dirac fermions. We show how one of these latter category of particles can contribute 
quite significantly to neutrinoless double beta decay. The neutrino virtuality momentum is 
found to play a crucial role in the non-standard contributions leading to the prediction 
of the pseudo-Dirac fermion mass in the range of $120\, \mbox{MeV}-500\, \mbox{MeV}$. 
When the Dirac neutrino mass matrix in the inverse seesaw formula is similar to the up-quark 
mass matrix, characteristic of high scale quark-lepton symmetric origin, the predicted 
branching ratios for lepton flavor violating decays are also found to be closer to the accessible 
range of ongoing experiments.

\end{abstract}
\maketitle
{\bf \small \hspace{-0.5cm} I. INTRODUCTION:}
The standard gauge theory of strong, weak, and electromagnetic interactions has confronted 
numerous experimental tests while the last piece of evidence on the Higgs boson is currently 
under rigorous scrutiny at the Large Hadron Collider (LHC). In spite of these, 
neutrino oscillation data uncovering tiny masses of left-handed (LH) neutrinos call for physics 
beyond the standard model (SM) which is most simply achieved via canonical seesaw mechanism \cite{seesaw1,valle-seesaw}
that requires the addition of one heavy right-handed (RH) neutrino per generation provided 
both LH and RH neutrinos are Majorana fermions \cite{Maj}. Several other forms of seesaw mechanism 
\cite{type-IImoh,seesaw2,seesaw-others} also require Majorana fermions. 
Quite interestingly, ongoing experiments on neutrinoless double beta decay ($0\nu \beta \beta$) 
\cite{expt-0nu2beta} is expected to resolve the issue between Majorana \cite{Maj} or Dirac \cite{Dirac} 
nature of the neutrino \footnote{Besides the two distinct possibilities, Dirac or Majorana, very recently a new 
hypothesis has been advanced in which neutrinos could be schizophrenic \cite{schezophenic}.}. 
In contrast to the predicted small contribution to the $0\nu\beta\beta$ decay rate in 
the SM, there has been quite significant, or even more dominant predictions if, at the TeV scale, 
there is left-right (LR) gauge theory \cite{tello-senj, mkp-spatra}. 
Even, attempts have been made to predict nonstandard contributions to $0\nu \beta \beta$ decay 
rate due to the mediation of pseudo-Dirac neutrinos where each of them is considered to 
be a pair of Majorana neutrinos \cite{reso-pseudo,mkp-spatra}. 
While the possibility of left-handed neutrinos being pseudo-Dirac has been shown to be highly 
challenging \cite{LHpseudoD}, contribution of a fourth generation heavy pseudo-Dirac neutrino to 
$0\nu \beta \beta$ has been explored with the condition that its mass should be greater than $M_Z/2$ 
\cite{HpseudoD}. If the Dirac neutrino mass matrix occurring 
in seesaw formulas has its left-right symmetric or quark-lepton symmetric origin, descending 
from Pati-Salam symmetry \cite{pati} or SO(10) grand unified theory \cite{so10-sm} at high scales, 
then the canonical seesaw scale is too large to be 
experimentally tested by high energy accelerators including LHC. Alternatively, the inverse seesaw 
mechanism \cite{inv,invo}, which requires one RH neutrino as well as an additional sterile fermion per generation, operates at TeV scale and is, 
therefore, experimentally verifiable. In this framework while the LH light neutrinos are Majorana fermions, 
the RH neutrinos are pseudo-Dirac by nature having heavier masses.

In this letter we show that the inverse seesaw formula explaining the light neutrino masses and mixings 
permits the lightest of the three pseudo-Dirac neutrinos in the mass range ($120-500$) MeV leading to 
new contributions to $0\nu \beta \beta$ decay comparable to, or much more than, those due to the 
exchanges of the light left-handed neutrinos. The neutrino virtuality momentum \cite{nuvirt-moh,nuvirt}, 
$|p| \sim 190$ MeV, is noted to play a crucial role in such new contributions. The origin of Dirac 
neutrino mass matrix is also found to be important in our estimations in predicting lepton flavor 
violating decays accessible to ongoing experimental searches. As our results are also applicable 
in the inverse seesaw extension of the minimal supersymmetric standard model (MSSM), they are 
consistent with gauge coupling unification at the MSSM-GUT scale, $M_U \simeq 2 \times 10^{16}$ GeV.

{\bf \small \hspace{-0.5cm} II. THE INVERSE SEESAW EXTENSION:}
As is customary to the implementation of inverse seesaw mechanism, we add two fermion singlets 
to each generation of the SM, with or without supersymmetry. While we call the first type of singlet a RH neutrino 
($N_R$), the second type of singlet is named as a sterile neutrino ($S_L$) and, in the 
($\nu_L$, $N^c_R$, $S_L$) basis, the $9 \times 9$ neutrino mass matrix is \cite{inv}
\begin{equation}
\mathcal{M}_\nu= \left( \begin{array}{ccc}
                0     & M_D   & 0   \\
              M^T_D   & 0     & M^T \\
              0       & M   & \mu_S
        \end{array} \right) \, ,
\label{eqn:numatrix}       
\end{equation}
where $M_D$ is the Dirac mass term of the neutrino, and $M$ is the heavy Dirac mass matrix relating 
$N_R$ and $S_L$. The matrices $M_D$ and $M$ are in general $3\times3$ complex in flavor space 
whereas the $\mu_S$ is $3\times3$ complex symmetric matrix.

Transformation from flavor to mass basis and diagonalization are achieved through
\begin{equation}
|\nu\rangle_f=\mathcal{V}^*|\nu\rangle_m\,,
\end{equation} 
\begin{equation}
\mathcal{V}^\dagger \mathcal{M}_\nu \mathcal{V}^* 
     =  \hat{\mathcal{M}}_\nu
= \text{Diag}\{m_{\nu_i}; {M_\zeta}_j \} \, ,
\end{equation} 
where $|\nu\rangle_m=(\tilde{\nu}_i, \zeta_j)^T$ represents the three light and six heavy mass states, and $i$ and $j$ run 
over the light and heavy mass eigenstates, respectively. With $\mu_S, M_D \ll M$, the matrix $\mathcal{M}_\nu$ can be 
block diagonalized to light and heavy sectors
\begin{eqnarray}
& &m_{\nu} \simeq \left(\frac{M_D}{M}\right)\, \mu_S\, \left(\frac{M_D}{ M}\right)^T \, , \nonumber \\
& &M_{H} \simeq\bmt
0&M^T\\
M&\mu_S
\emt \, .
\label{eqn:numass}
\end{eqnarray}
where $m_\nu$ has the well known inverse seesaw formula \cite{inv} and $M_H$ is the mass matrix for 
heavy pseudo-Dirac pairs of comparable masses with splitting of the order of $\mu_S$. The $\mu_S$ term 
in the Lagrangian breaks the leptonic global symmetry, $U(1)_L$, which is otherwise preserved in the 
standard model in the limit $\mu_S\rightarrow 0$ rendering all the LH neutrinos to be massless. 
Hence the small $\mu_S$ should be a natural parameter in this theory in the 't Hooft sense \cite{thooft}. The above 
block diagonalized matrices are further diagonalized through the ${\rm PMNS}$ matrix, $U_\nu$, and
a $6\times 6$ unitary matrix $U_H$, respectively, so that
\begin{eqnarray}
\mathcal{V}\simeq
\bmt
1-\frac{1}{2}B^*B^T&B^*\\
-B^T&1-\frac{1}{2}B^TB^*
\emt
\bmt
U_{\nu}&0\\
0&U_{H}
\emt\,,
\end{eqnarray}
where
\begin{eqnarray}
B^T\simeq
\bmt
-{M^*}^{-1}\mu^*_S (M_D{M}^{-1})^\dagger\\
(M_D{M}^{-1})^\dagger
\emt\simeq
\bmt
0\\
X^\dagger
\emt\,.
\end{eqnarray}
Hence, in the leading order approximation, $\mathcal{V}$ can be written as
\begin{eqnarray}
\mathcal{V}
\simeq
\bmt
1-\frac{1}{2}XX^\dagger &0&X\\
0&1&0\\
-X^\dagger &0&1-\frac{1}{2}X^\dagger X
\emt
\bmt
U_{\nu}&0\\
0&U_{H}
\emt,
\end{eqnarray}
where $X=(M_DM^{-1})$, and all the elements in the first block are $3\times 3$ matrices.

\begin{table}[htb!]
\centering
\begin{tabular}{c|c}
\hline
{\small \bf Neutrino oscillation parameters}  & {\bf \small Globally fitted values} \\[0.1in] 
\hline \hline
$\Delta m^2_{\rm {sol}}   [\mbox{eV}^2]$   & $7.58 \times 10^{-5}$  \\[0.05in] 
$|\Delta m^2_{\rm {atm}}| [\mbox{eV}^2]$ & $2.35 \times 10^{-3} $ \\[0.0in]
$\sin^2 \theta_{12}$            & 0.320  \\
$\sin^2 \theta_{23}$            & 0.427  \\
$\sin^2 \theta_{13}$            & 0.0246 \\[0.05in] 
$\delta_{\rm CP}$                   & 0.8 $\pi$ \\[0.05in] 
\hline \hline
\end{tabular}
\caption{Mass squared differences, mixing angles, and $CP$-phase from global fits 
         to neutrino oscillation data \cite{DAYA-BAY}.}
\label{tab:input-param}
\end{table}

{\bf \small \hspace{-0.45cm} (II. A) ${\bf \mu_S}$ from neutrino oscillation data:}
The inverse seesaw formula in eqn.~(\ref{eqn:numass}) predicts light neutrino mass matrix 
in terms of three other matrices, $M_D$, $M$, and $\mu_S$. At first we take $M_D \simeq M_{\ell}$, 
the charged lepton mass matrix, which may arise if the SM originates from high scale 
left-right gauge symmetry, $SU(2)_L\times SU(2)_R\times U(1)_{B-L}\times SU(3)_C\stackrel {M_R}{\longrightarrow}SM$, 
where $M_R>>M_W$. Assuming the matrix $M$ to be diagonal for the sake of simplicity and 
using $M_D = \mbox{diag} \{m_e, m_\mu, m_\tau \} = \{0.0005, 0.1, 1.7 \}$ GeV, we obtain $\mu_S$ from  
global fits to the neutrino oscillation data \cite{DAYA-BAY} given in TABLE \ref{tab:input-param}
\begin{widetext}
\begin{eqnarray}
\mu_S\, (\mbox{GeV}) &=& X^{-1}\, 
\mathcal{N} \hat{m}_\nu \mathcal{N}^T\, {X^T}^{-1} 
\label{eqn:mus-matrix} \\
& &\hspace{-0.5cm}={ \bf 
\left(
\begin{array}{ccc}
 6.71\times 10^{-7} +1.96 \times 10^{-7}\, {\bf i} & 
 -1.17\times 10^{-8}-3.22 \times 10^{-8}\, {\bf i} & 
 -3.71\times 10^{-8} -2.03 \times 10^{-8}\, {\bf i}   \\
 -1.17\times 10^{-8}-3.22 \times 10^{-8}\, {\bf i} &
  1.53\times 10^{-08}-2.22 \times 10^{-10}\, {\bf i} & 
  7.0\times 10^{-9} -2.83 \times 10^{-9}\, {\bf i} \\
   -3.71\times 10^{-8} -2.03 \times 10^{-8}\, {\bf i} &
  7.0\times 10^{-9} -2.83 \times 10^{-9}\, {\bf i} & 
-5.50\times 10^{-9}+5.26 \times 10^{-11}\, {\bf i}
\end{array}
\right),
\label{eqn:mus-value}
} 
\end{eqnarray}
\end{widetext}
where {\small \bf $\mathcal{N}=(1-\eta)\,U_{\nu}$} and {\small \bf $\eta=\frac{1}{2} 
XX^\dagger$} is a measure of unitarity violation. This particular structure of $\mu_S$ 
has been derived using, as an example, the normal hierarchical (NH) light neutrino masses $\hat{m}_{\nu}^{\rm diag} 
= {\rm diag}(\text{0.00127 ~eV}, ~\text{0.00885 ~eV}, ~\text{0.0495 ~eV})$ and non-degenerate 
eigenvalues of $M = \mbox{diag}\, \{0.2, 2.6, 23.7 \}$ GeV. Similar analysis predicts somewhat 
different structures of $\mu_S$ for inverted hierarchical (IH) and quasi-degenerate (QD) pattern 
of the light neutrinos and can further be easily obtained for degenerate $M_{1} = 
M_{2} = M_{3}$ or,  partially-degenerate $M_{1} = M_{2} \ll M_{3}$ 
after taking care of the phenomenological bounds $|\eta_{ee}|  < 2.0 \times 10^{-3}$, 
$|\eta_{\mu \mu}| < 8.0\times 10^{-4}$, 
and $|\eta_{\tau \tau}| < 2.7 \times 10^{-3}$. Our ansatz with $M={\rm diag}(M_1, M_2, M_3)$ gives 
$$\eta = \mbox{diag} \left(\frac{1.25 \times 10^{-7}}{M^2_1}, \, 
\frac{0.005}{M^2_2}\, , \frac{1.35}{M^2_3} \right) \, ,$$
where all masses on the right hand side are in $\mbox{GeV}$.

\begin{figure}[htb!]
\centering
\includegraphics[width=0.48\textwidth]{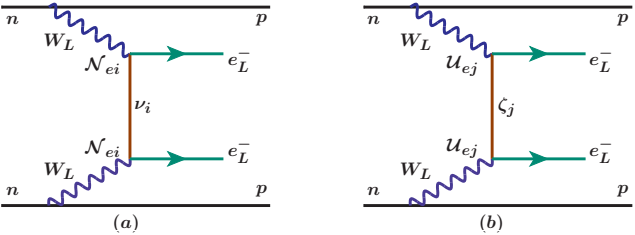}
\caption{Feynman diagrams contributing to neutrinoless double beta decay  
         due to light neutrino exchanges (left-panel) and heavy pseudo-Dirac neutrino exchanges (right-panel).} 
\label{fig:feyndia}
\end{figure}

{\bf \small \hspace{-0.45cm}III. NEUTRINOLESS DOUBLE BETA DECAY PREDICTIONS}
Two separate contributions due to light and heavy neutrino exchanges to $0\nu \beta \beta$ transition 
become transparent by writing the flavor eigenstates as linear combination of light and heavy mass 
eigenstates 
$$\nu_{\alpha}= {\cal N}_{\alpha i}\nu_i + {\cal U}_{\alpha j} \zeta_j \, ,$$ 
where ${\cal U}\simeq(0, X)U_H$ is a $3\times 6$ matrix. Then the weak charge-current Lagrangian can be expressed as 
\begin{eqnarray}
	{\cal L}_{\rm CC} &=& \frac{g}{\sqrt 2}
	W^\mu_L \bar{\ell}_\alpha\gamma^\mu P_L \nu_\alpha + {\rm h.c.} \nonumber \\
	& = & \frac{g}{\sqrt 2} W^\mu_L \bar{\ell}_\alpha 
	\gamma^\mu P_L ({\cal N}_{\alpha i}\nu_i + {\cal U}_{\alpha j} \zeta_j)
	+ {\rm h.c.} \, ,
\label{eq:12} 
\end{eqnarray}
resulting in two different categories of Feynman amplitudes:
\begin{itemize}
\item $\mathcal{A}^{\nu}_{LL}$ which arises from the Feynman diagram of 
Fig.~\ref{fig:feyndia}(a) due to only light neutrino exchanges
\begin{eqnarray}
& & \mathcal{A}^{\nu}_{LL} = G^2_F\, \mathcal{N}^2_{ei}\, \frac{{m_\nu}_i}{p^2} \, ,
\end{eqnarray}
where $\langle p \rangle \simeq 190$~MeV represents neutrino virtuality momentum \cite{nuvirt-moh,nuvirt}.
 
\item $\mathcal{A}^{\zeta}_{LL}$ which arises from the Feynman diagram of 
Fig.~\ref{fig:feyndia}(b) due to heavy pseudo-Dirac neutrinos,
\begin{eqnarray}
& & \mathcal{A}^{\zeta}_{LL}   = G^2_F\, \left(\mathcal{U} \right)^2_{ej}\, \frac{{M_\zeta}_j}{p^2- {M^2_\zeta}_j} \, .
\end{eqnarray}
\end{itemize}

The mass eigenstates of heavy pseudo-Dirac neutrinos are $\left(\zeta^+_1,\zeta^+_2,\zeta^+_3;\zeta^-_1, 
\zeta^-_2,\zeta^-_3 \right)$ with almost degenerate pairs ($\zeta^+_{k},\zeta^-_{k}$;\, k=1,2,3) but having  
small mass difference $\mu_S$ between the members of the pair and the flavor states are $\left(N_{1}, 
N_{2}, N_{3}; S_{1}, S_{2}, S_{3}\right)$. The mixing matrix for these pseudo-Dirac neutrinos 
have been discussed in Sec-{\bf II}. The half-life of $0\nu \beta \beta$ transition is then found 
to be 
\begin{eqnarray}
\bigg[T^{0\nu \beta \beta}_{1/2} \bigg]^{-1} 
= \mathcal{K}_{0\nu} \bigg[ 
\left|m^{ee}_{\nu,{\rm \bf \small LL}} + M^{ee}_{\zeta,{\rm \bf \small LL}}\right|^2
\bigg] \, ,
\end{eqnarray}
where $\mathcal{K}_{0\nu}$ contains phase space factors plus nuclear matrix elements and 
{\bf $m^{ee}_{\nu,{\rm \bf \small LL}}$} ({\bf $M^{ee}_{\zeta,{\rm \bf \small LL}}$}) represents the 
effective neutrino mass derived from light neutrino (heavy pseudo-Dirac neutrino) exchanges in the 
mass basis. The analytic forms of the two effective masses have been estimated for this model as shown 
in TABLE. {\ref{tab:eemass}}:
\begin{table}[htb]
\begin{center}
\begin{tabular}{|c|c|}
\hline
{\bf Effective mass}     &  {\bf Analytical expression } 
\rule{0pt}{2.1em}\\  [10pt]
\hline \hline
{\bf $m^{ee}_{\nu,{\rm \bf \small LL}}$}  &  
 $ {\cal N}_{ei}^2\, m_{\nu_i}$
  \rule{0pt}{2.1em}\\  [10pt]
{\bf $M^{ee}_{\zeta,{\rm \bf \small LL}}$}  &  
 $\left(\mathcal{U}_{ej}\right)^2\frac{M_{\zeta_j}}{p^2-M^2_{\zeta_j}} |\langle p\rangle|^2$
  \rule{0pt}{2.1em}\\  [10pt]
\hline \hline
\end{tabular}
\end{center}
\caption{Effective mass parameter for standard (non-standard) contributions due to light (heavy 
pseudo-Dirac) neutrino exchanges for $0\nu\beta\beta$ decay.}
\label{tab:eemass}
\end{table}

\begin{figure*}[htb]
\includegraphics[scale=0.33,angle=-90]{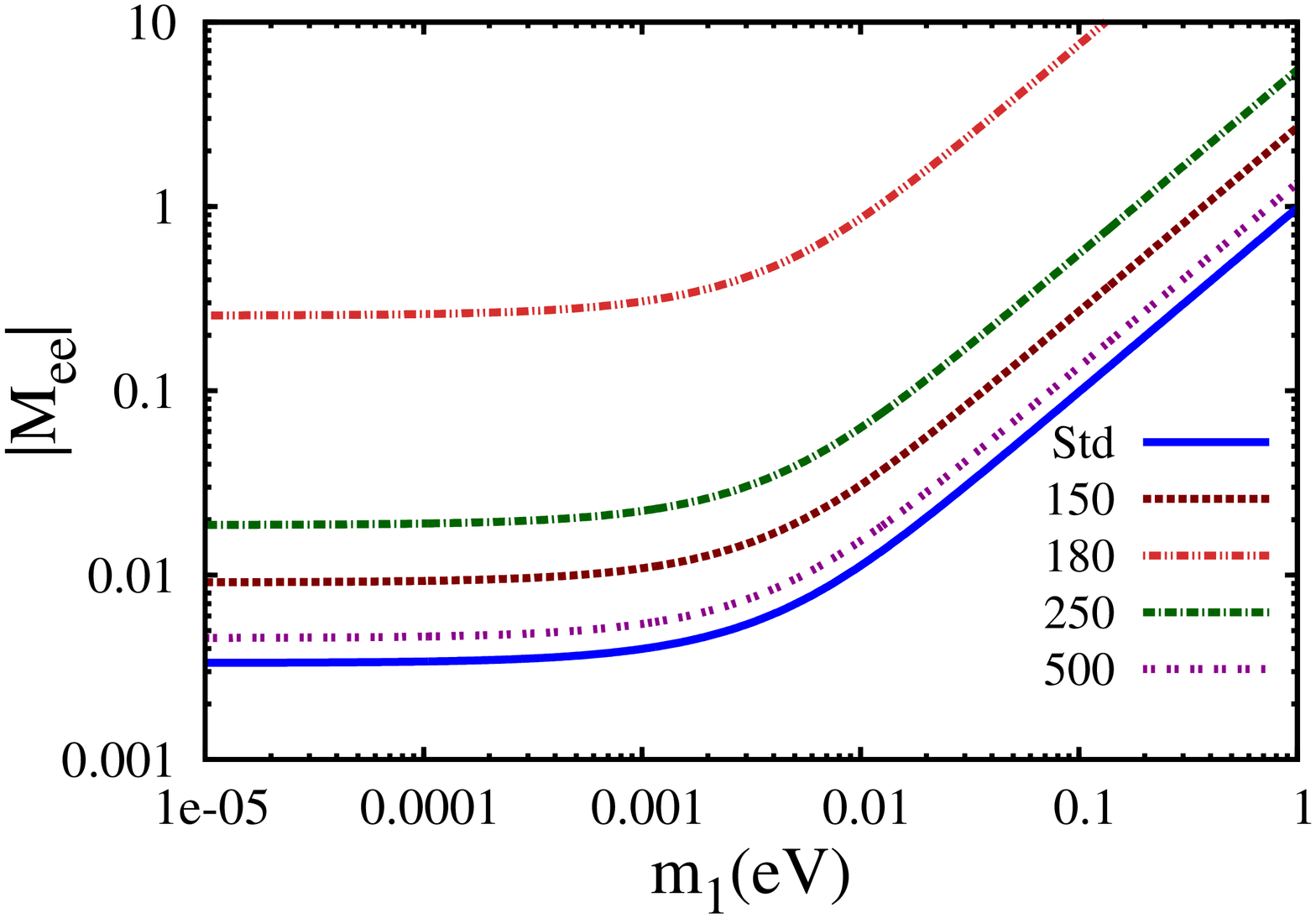}
 \includegraphics[scale=0.33,angle=-90]{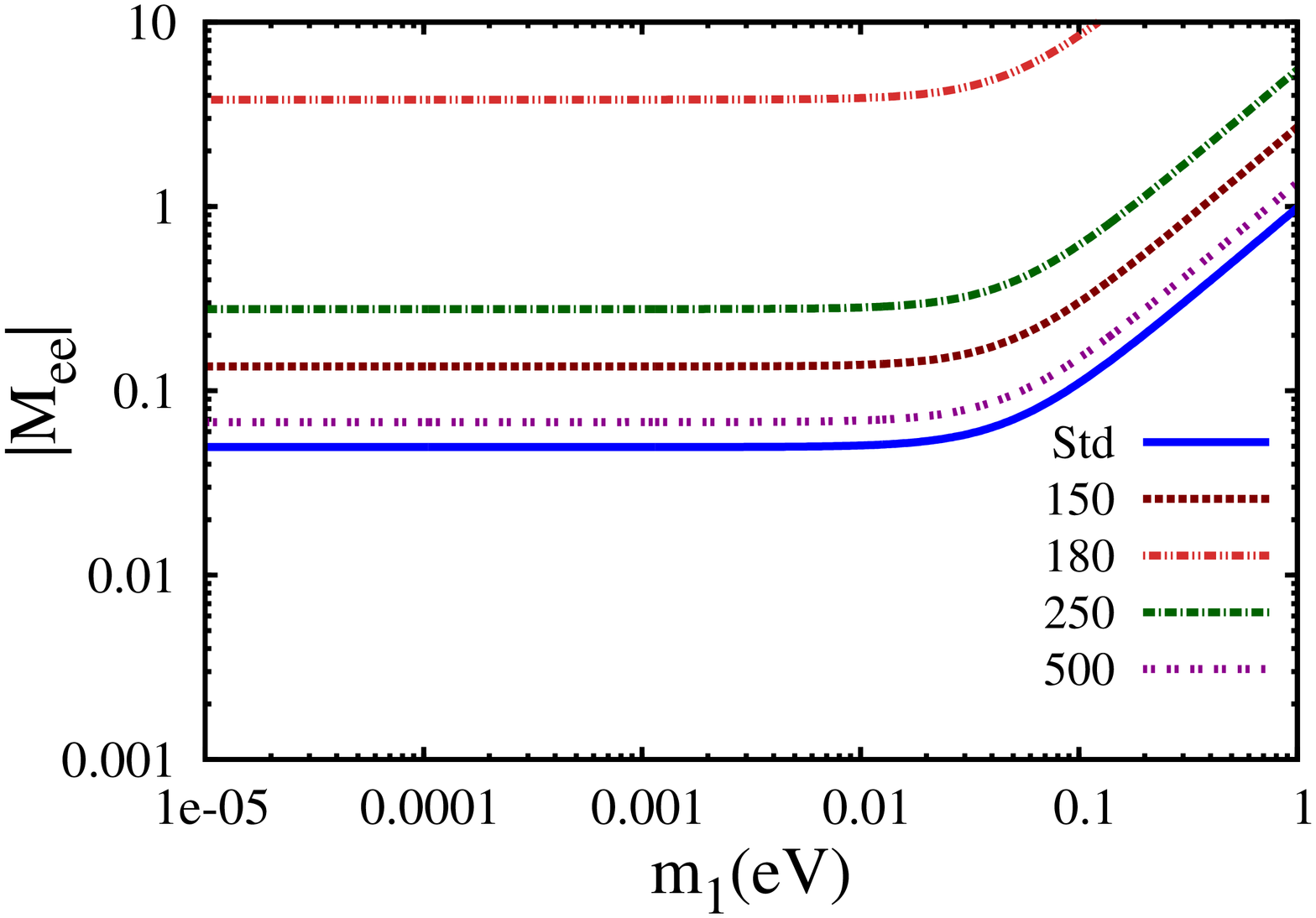}
\caption{Predictions of effective mass $|M_{ee}|$ in $0\nu\beta\beta$ decay with Dirac neutrino mass $M_D=$ charged
lepton mass $M_l$ and the diagonal structure of $M$ for NH (IH) pattern of light LH neutrino masses as shown in 
the {\it left (right) panel}. The standard contribution is shown by solid line and nonstandard contributions with
pseudo-Dirac neutrino exchanges of different masses expressed in MeV are shown by other lines.}
\label{plot:sm-pseudo-a}
\end{figure*}

\begin{figure*}[htb]
\includegraphics[scale=0.33,angle=-90]{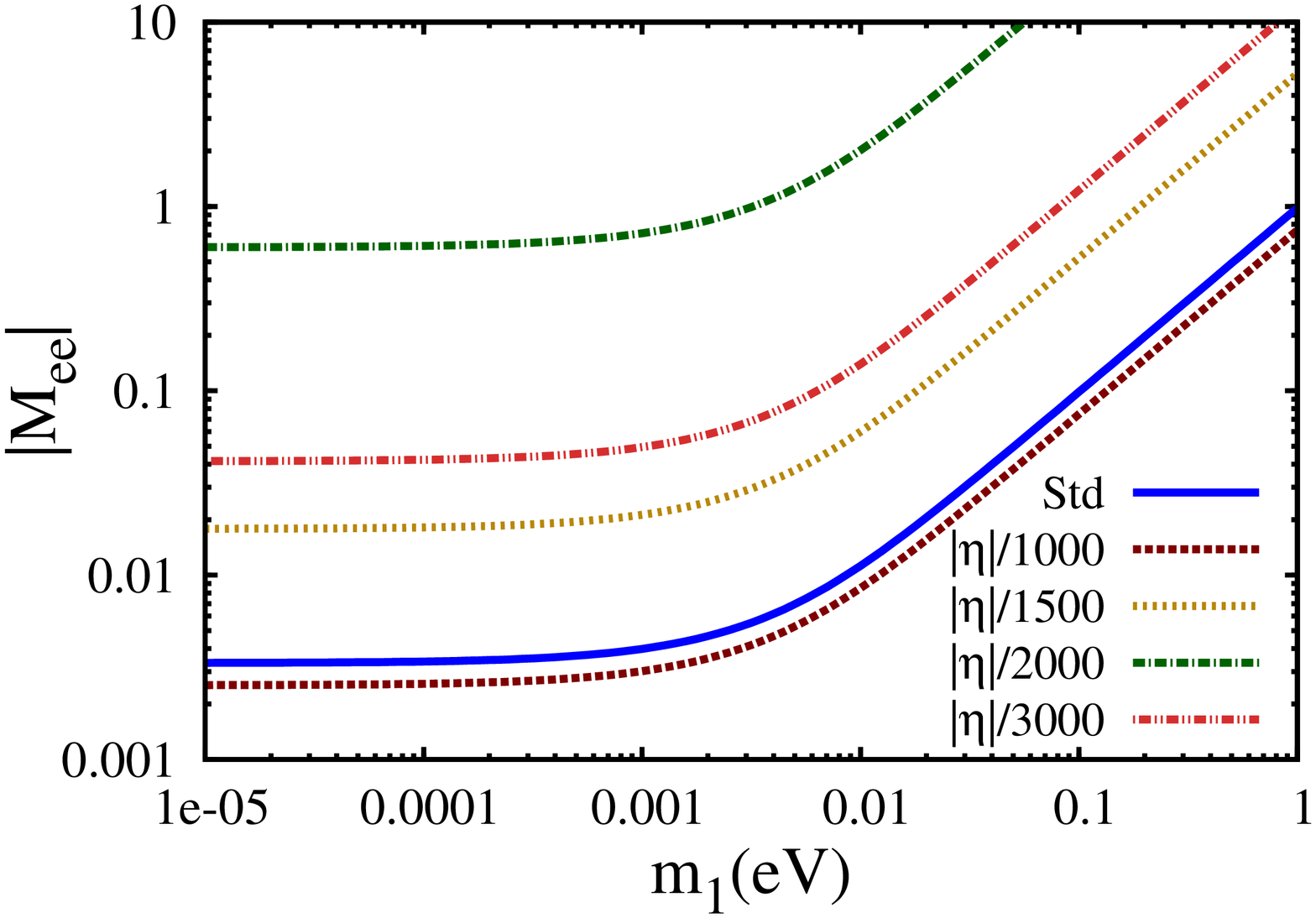}
\includegraphics[scale=0.33,angle=-90]{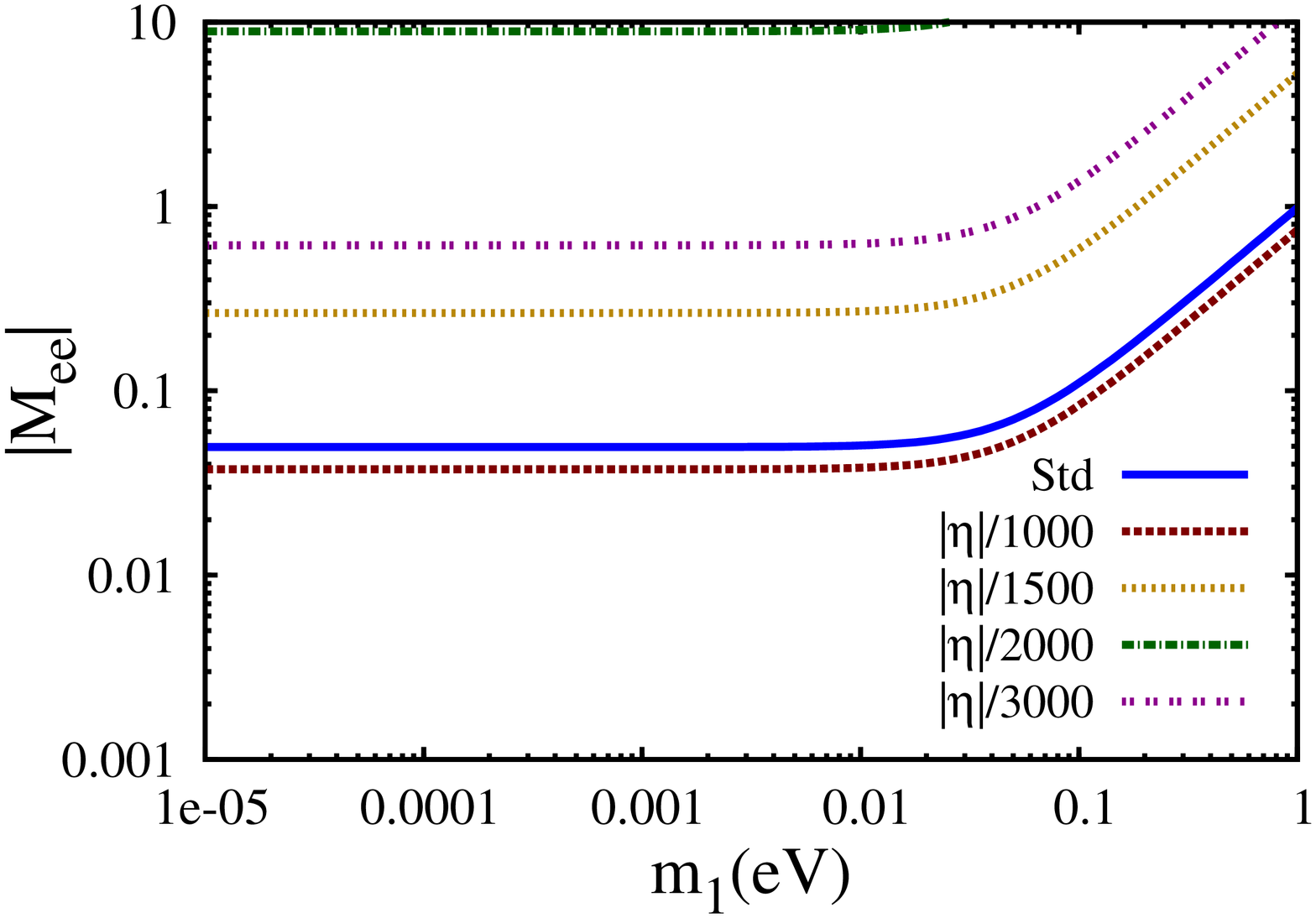}
\caption{Same as Fig. \ref{plot:sm-pseudo-a} but now with non-diagonal structure of $M$ and reduced values of nonunitarity
matrices $\eta$ as described in the text.}
 \label{plot:sm-pseudo-b}
\end{figure*}

\begin{figure*}[htb]
\includegraphics[scale=0.33,angle=-90]{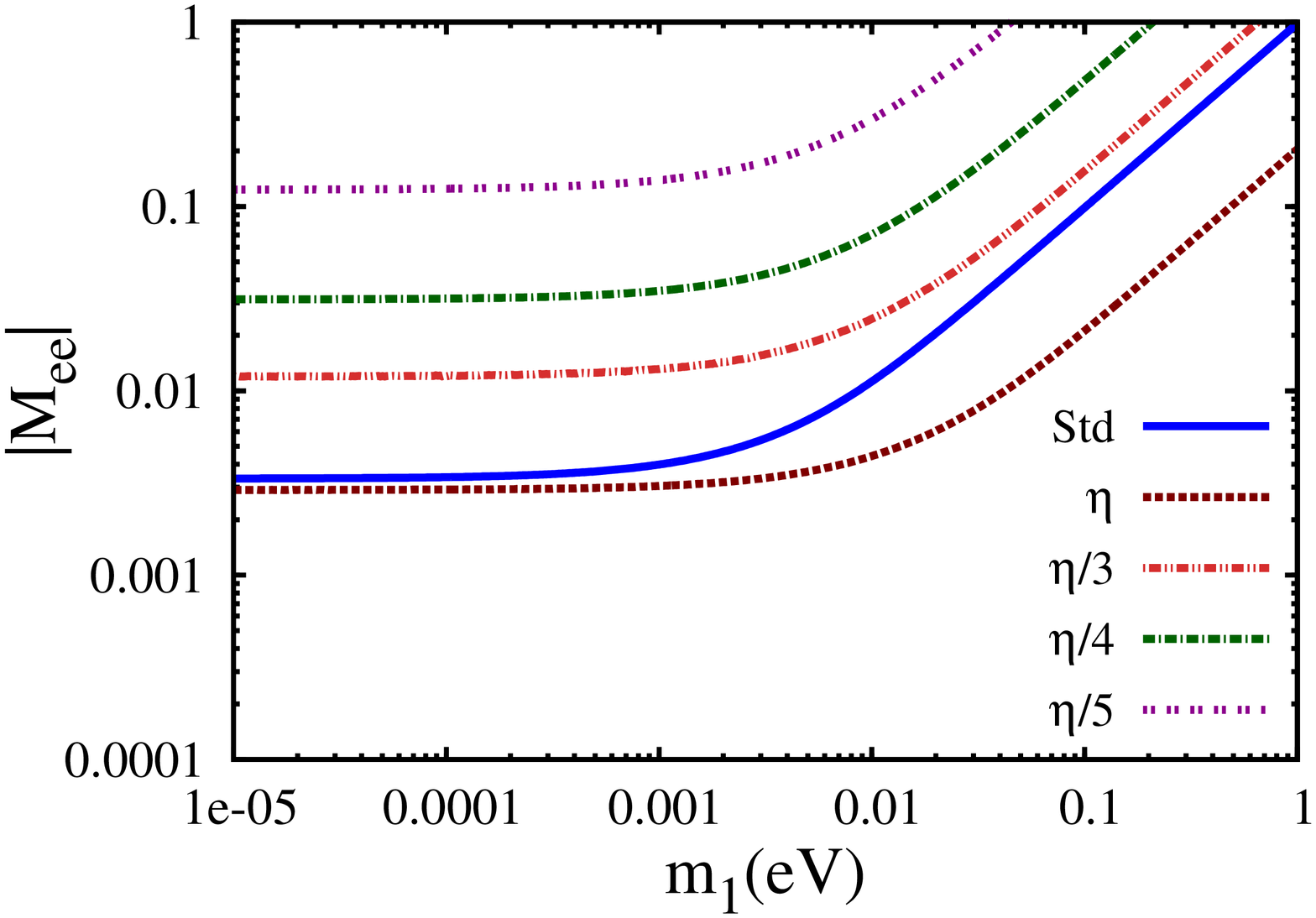}
 \includegraphics[scale=0.33,angle=-90]{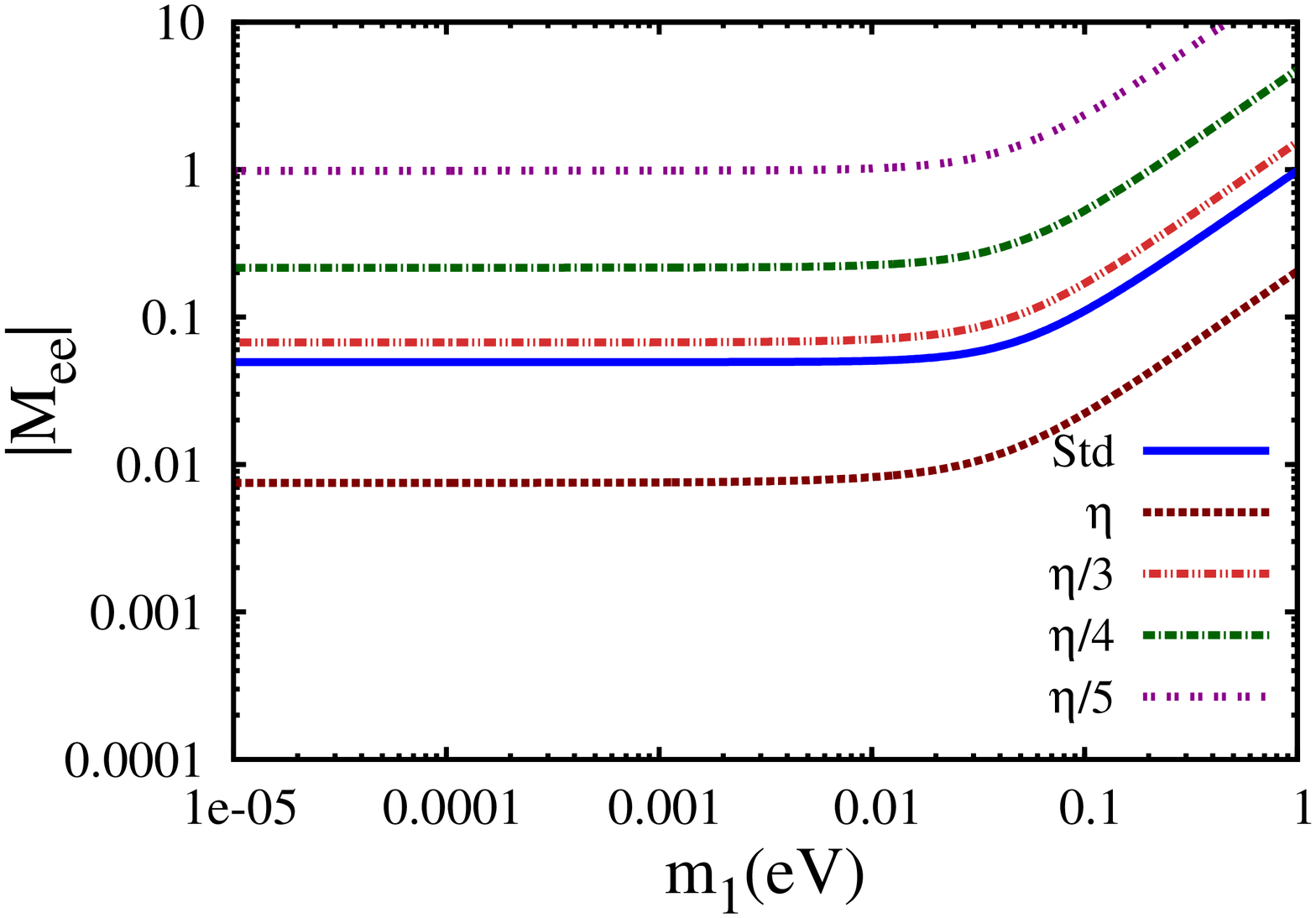}
\caption{The effective mass parameter, $|M_{ee}|$, prediction for $0\nu\beta\beta$ decay due to light (Solid line) and 
pseudo-Dirac (dashed lines) neutrino exchange where $M_D\simeq M_u$ and $\eta=|\eta_{max}|/n,\, n=1,3,4,5$ as discussed in the text.}
\label{plot:sm-pseudo-c}
\end{figure*}

We discuss below three different cases:

{\bf \hspace*{-0.45cm} (III. A) The standard contribution.}
It is well known that the standard contributions due to light neutrino exchanges are dependent 
on their allowed mass patterns; {\it normal hierarchical} (NH), {\it inverted hierarchical} (IH), 
or {\it quasi-degenerate} (QD),
\bea \label{eq:mee}
\nonumber m^{\nu}_{\rm ee,{\bf LL}}  \simeq U^2_{e1}\, m_{\nu_1} + U^2_{e2}\, e^{2 i \alpha} \, m_2 &+ U^2_{e3}\, e^{2 i \beta} \, m_3 
 \\[0.3cm]
\Rightarrow |m^{\nu}_{\rm ee,{\bf LL}}| \simeq \left\{
\baz 
0.004\, \mbox{eV}
& \mbox{ NH,} \\[0.2cm]
0.048\, \mbox{eV}
& \mbox{ IH,} \\[0.2cm]
0.1\, \mbox{eV} 
& \mbox{ QD.} 
\ea \right. 
\eea
In our case, $\mathcal{N}_{ei} \simeq U_{ei}$ and light neutrino exchanges in the mass 
basis gives almost the same contributions which are presented by solid lines 
shown in Fig. \ref{plot:sm-pseudo-a}, Fig. \ref{plot:sm-pseudo-b}, and Fig. \ref{plot:sm-pseudo-c}. 

{\bf \hspace*{-0.45cm} (III. B) $M_{\zeta_j} \gg |p|$:~}
In the inverse seesaw extension under study, in addition to the standard effective mass 
parameter, the additional effective 
mass parameter for $|M_{\zeta_j}| \gg |p|$ satisfies $\langle M_{\zeta_{\pm}} \rangle^{-1} = 
\left(\mathcal{U}^{\pm}\right)^2\frac{-1} {M_{\zeta_{\pm}}}$. This results in new 
contribution to $0\nu \beta \beta$ transition half-life 
\begin{eqnarray}
\bigg[T^{0\nu \beta \beta}_{1/2} \bigg]^{-1} 
&=& \mathcal{K}_{0\nu} 
\left|\, |\langle p\rangle|^2 \left( \frac{1}{\langle M_{\zeta_{+}} \rangle} - 
\frac{1}{\langle M_{\zeta_{-}} \rangle}
\right) \right|^2 \nonumber \\
&\simeq& \mathcal{K}_{0\nu} 
\left|\, |\langle p\rangle|^2\,\left(\mathcal{U}^{\pm}\right)^2_{ek}\, \frac{{\mu_S}_{kk}}{M^2_{kk}} \right|^2 \, ,
\end{eqnarray} 
where ${\mu_S}_{kk}$ and $M_{kk}$ are the eigenvalues of $\mu_S$ and $M$, respectively. 
One example of this case has been shown in Fig. \ref{plot:sm-pseudo-a} for $M_{\zeta_1} = 
0.5$ GeV where the predicted effective mass parameter is nearly 3/2 (4) times larger 
than the standard prediction for NH (IH) case.

{\bf \hspace*{-0.45cm} (III. C) $M_{\zeta_j} \simeq |p|$:~}
In this region where different allowed values of $M_{\zeta_{j}}$ are of the order of neutrino 
virtuality momentum $|p| \simeq 190$ MeV, the new contribution to neutrinoless double beta decay due to 
heavy pseudo-Dirac neutrino exchange is found to be more dominant than the standard contribution 
and the $0\nu \beta \beta$ transition half-life is given below
\begin{eqnarray}
{\tiny \bigg[T^{0\nu \beta \beta}_{1/2} \bigg]^{-1}_{\rm pseudo-Dirac}} 
\hspace*{-1.5cm}&=& { \tiny \mathcal{K}_{0\nu} 
\left|\, |\langle p\rangle|^2\, \mathcal{U}^{\pm}\, 
\left( \frac{M_{\zeta_{+}}}{p^2 -M^2_{\zeta_{+}}} - \frac{M_{\zeta_{-}}}{p^2 -M^2_{\zeta_{-}}}
\right) \right|^2 } \nonumber \\
&\simeq& 
\mathcal{K}_{0\nu}
\left|\, |\langle p\rangle|^2\,\left(\mathcal{U}^{\pm}\right)^2_{ek}\, 
\frac{{\mu_S}_{kk}}{p^2 - M^2_{kk}} \right|^2 \, .
\end{eqnarray}
The predicted new values of the effective mass parameters arising solely due to pseudo-Dirac 
neutrino exchanges have been shown in Fig. \ref{plot:sm-pseudo-a} in the {\it left-pannel} 
({\it right-pannel}) for NH (IH) patterns of the light neutrino masses, respectively, where 
$M_{\zeta_1}=(0.15-0.5)$ GeV. It is quite clear from the plots that even for 
$M_{\zeta_{1}}= 0.25$ GeV or, $0.5$ GeV, the new contributions are 3-6 times 
larger than the standard ones. While for the value of $M_{\zeta_{1}}= 0.18$ GeV, 
the contribution is nearly 100 times larger shown in Fig. \ref{plot:sm-pseudo-a}. 
This large enhancement occurs as $M_{\zeta_1}$ approaches the vicinity of the neutrino 
virtuality momentum, $|p| \simeq 190$ MeV. We point out that such important effects of pseudo-Dirac 
neutrino masses are found for the first time in this work.

{\bf \small \hspace{-0.5cm} IV. LEPTON FLAVOR VIOLATION WITH DOMINANT $\bm{0\nu\beta\beta}$ DECAY RATE:}
We have clearly shown that the predicted non-standard contributions to neutrinoless double beta 
decay rate are dominant for the lightest allowed pseudo-Dirac neutrino mass $M_{\zeta_1} \simeq 
(0.15-0.5)$ GeV. However, because of the diagonal nature of $M_D$ and assumed structure of $M$, 
the branching ratios for lepton flavor violating (LFV) decays, $\mu \rightarrow e + \gamma$, $\tau 
\rightarrow e + \gamma$, and $\tau \rightarrow \mu + \gamma$ are as small as the SM predictions. 
In the next two examples we adopt plausible parametrization predicting 
significantly larger contribution to these branching ratios while retaining the dominant contributions 
to $0\nu \beta \beta$ transition. 
 
{\bf \hspace{-0.5cm} (IV. A) $M_D \simeq M_{\ell}$ with non-diagonal $M$:~} 
We generate non-diagonal matrix $M$ to satisfy the existing phenomenological bounds on the elements of $\eta$ 
\cite{antu-eta},
\begin{eqnarray} 
\label{eq:exptbound-eta}
& &|\eta_{ee}|  < 2.0 \times 10^{-3}, \quad |\eta_{e \mu}| < 3.5\times 10^{-5}\, , \nonumber \\
& &|\eta_{e \tau}| < 8.0\times 10^{-3}, \quad |\eta_{\mu \mu}| < 8.0\times 10^{-4} \, , \nonumber \\
& &|\eta_{\mu \tau}| <  5.1\times 10^{-3} , \quad |\eta_{\tau \tau}| < 2.7 \times 10^{-3} \, .
\end{eqnarray}
Using the parametrization of the type used in ref. \cite{CI-para}, $M$ can be expressed as
\begin{eqnarray}
\label{eq:CIpara-M}
 \left[M \left( \dfrac{1}{\sqrt{2 \eta}} O^T M_D \right)^{-1} \right]^\dagger & 
\left[M \left(\dfrac{1}{\sqrt{2 \eta}} O^T M_D \right)^{-1} \right]   \nonumber \\ 
={\bf 1}_{3} = V^\dagger \, V
\end{eqnarray}
where $O$ is the matrix diagonalizing $|\eta|$ and $V$ is an arbitrary unitary matrix. 
Choosing, for the sake of simplicity, $V =1_{3}$ and we note that that a lightest pair 
with ${M_\zeta}_1\simeq 0.16$~GeV, in the vicinity of neutrino virtuality momentum, 
can be achieved by suitable rescaling, {\it e.g.} $\eta_{\alpha \beta} \to \eta_{\alpha \beta}/(1500)$. 
After this scaling we find 
\begin{eqnarray}
M(\mbox{GeV})&=& \left(\frac{1}{\sqrt{2 \eta}} O^T M_D \right) \\
&=&
\bmt 
0.092 i& 14.08 i & -383.7 i \\
0.217& -70.36& -80.39\\
0.074& 8.866  & 320.5
\emt  \, , \nonumber
\label{eq:M}
\end{eqnarray}
where
\begin{eqnarray*} 
\label{eq:ortho-eta}
O = 
\left( \begin{array}{ccc} 
0.5874 i & 0.5446 & 0.5987 \\
0.4284 i & -0.8368 & 0.3409 \\
-0.6866 i & -0.0562 & 0.7248
\end{array} \right)\,.
\end{eqnarray*}
With the allowed mass eigenvalues for the heavy pseudo-Dirac neutrinos, $M_{\zeta} = 
\mbox{diag} \{0.159, 72.0, 506.4\}$ GeV, the predicted branching ratios for 
lepton flavor violating decays are \cite{lfv-illakovac}
\begin{eqnarray}
& &\text{Br}\left(\mu \rightarrow e + \gamma \right) = 1.56 \times 10^{-26} \, ,
\nonumber \\
& &\text{Br}\left(\tau \rightarrow e + \gamma \right) = 5.79 \times 10^{-27} \, ,
\nonumber \\
& &\text{Br}\left(\tau \rightarrow \mu + \gamma \right) = 1.10 \times 10^{-18} \, .
\end{eqnarray}
Although all the three branching ratios are much smaller than their corresponding 
experimental upper limits \cite{LFV}, they are considerably larger than the SM predictions.
However we note below that with $M_D$ similar to $M_u$, the up-quark mass matrix, a
phenomenon underlying the possible origin of SM from Pati-Salam \cite{pati} or SO(10) model, LFV
decays have much larger predicted values, accessible to ongoing experimental searches, while
similar predictions on dominant $0\nu\beta\beta$ decay are maintained.

\noindent
{\bf (IV. B) $M_D \simeq M_{u}$ and GUT connection:~}  In this case Dirac neutrino mass matrix is 
approximated to be up-quark mass matrix, which originates if the high scale symmetry is Pati-Salam or SO(10) GUT,
$SU(2)_L\times SU(2)_R\times SU(4)_C$ or SO(10) $\stackrel {M_R}{\longrightarrow}SM$. Using 
running masses $(m_u, m_c, m_t)=(0.00233, 1.275, 160)$~GeV and Cabbibo-Kobayashi-Maskawa mixing matrix, $V_{CKM}$ \cite{PDG},
\bea 
M_D(\rm GeV)\simeq M_u=V_{CKM}\hat{M}_uV^T_{CKM}\hspace{2.5cm}&& \nonumber\\
\hspace*{-0.4cm}=\bmt
0.067-0.004i&0.302-0.022i&0.55-0.53i\\
0.302-0.022i&1.48-0.0i&6.534-0.001i\\
0.55-0.53i&6.534-0.0009i&159.72+0.0i
\emt .
\eea
At first using the phenomenological bounds from eqn. (\ref{eq:exptbound-eta}) and saturating our ansatz for
$\eta_{\alpha\beta}=|\eta|_{max}/n$, $n=1,..5\,$, we search for matrix $M$ through eqn. (\ref{eq:CIpara-M}) which gives
$\mu_S$  from eqn. (\ref{eqn:mus-matrix}). 
We obtain for $n=4$
\bea 
M(\rm GeV)\hspace{6.5cm}&&\nonumber\\
\hspace*{-0.4cm}=\bmt
-5.9-3.45 i&0.2-60.72 i&5.15-1760 i\\
-10.44+2.13 i&-60.9-0.5 i&-598.0-12.16 i\\
7.08-5.09 i&70.86-0.18 i&1547-4.15 i
\emt \, , \nonumber \\
&&
\eea
$\mu_S(\rm eV)$
\bea= 
\bmt
-3.42-0.51 i & -1.92+5.55 i & 0.28-1.92 i\\
-1.92+5.55 i &39.1+5.68 i & -12.1+0.11 i\\
 0.28-1.92 i & -12.1+0.11 i & 4.10-0.68 i
\emt.
\eea

Our predictions on numerical values of the effective mass parameter for $0\nu\beta\beta$ are 
shown in Fig.~\ref{plot:sm-pseudo-c} for NH, IH and QD cases. For NH light neutrinos we find that the predicted value of $|M_{ee}|$ is increased by a factor 3 for $\eta=|\eta|_{max}/3$, corresponding to lightest pair ${M_\zeta}_1=131$~MeV, while the increment is 10 times for $\eta=|\eta|_{max}/4$ and ${M_\zeta}_1=152$~MeV, and 30 times for $\eta=|\eta|_{max}/5$ and ${M_\zeta}_1=169$~MeV. We find that the enhancement survives as long
as lightest pair ${M_\zeta}_1\simeq 120-350$~MeV. For the IH light neutrino masses the results are 
similar as shown on the right panel of Fig.~\ref{plot:sm-pseudo-c}. The branching ratios for 
lepton flavor violating decays predicted in this scenario with ${M_\zeta}=(0.152,39.5,2426)$~GeV are
\begin{eqnarray}
\label{eqn:branch_pred}
& &\text{Br}\left(\mu \rightarrow e + \gamma \right) = 3.6 \times 10^{-13} \, ,
\nonumber \\
& &\text{Br}\left(\tau \rightarrow e + \gamma \right) = 4.2 \times 10^{-14} \, ,
\nonumber \\
& &\text{Br}\left(\tau \rightarrow \mu + \gamma \right) = 3.3 \times 10^{-12}\, , 
\end{eqnarray}
while the present experimental limits at 90$\%$ C.L. on these branching ratios are $\text{Br} 
\left(\mu \rightarrow e + \gamma \right) \leq 1.2 \times 10^{-11}$, 
$\text{Br}\left(\tau \rightarrow e + \gamma \right) \leq 3.3 \times 10^{-8}$, and 
$\text{Br}\left(\tau \rightarrow \mu + \gamma \right) \leq 4.4 \times 10^{-8}$ \cite{LFV}. 
The projected reach of sensitivity in the future is  
$\text{Br}\left(\tau \rightarrow e + \gamma \right), ~ \text{Br}\left(\tau \rightarrow 
\mu + \gamma \right) \leq 10^{-9} $ and specifically $\text{Br}\left(\mu \rightarrow e 
+ \gamma \right) \leq 10^{-19}$ \cite{LFV}.

\begin{figure}[htb!]
\centering
\includegraphics[width=0.48\textwidth]{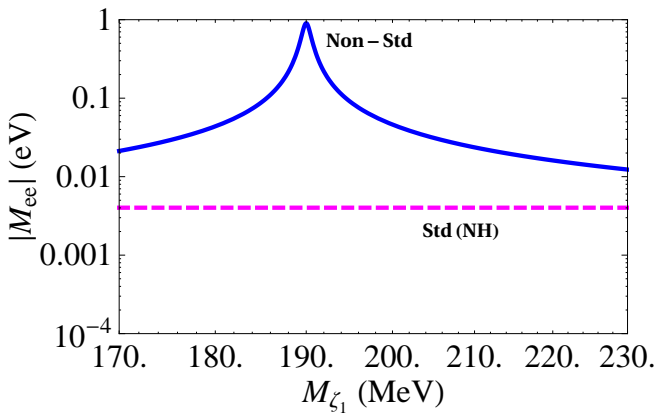}
\caption{Variation of effective mass $|M_{ee}|$ prediction as a function of lightest pseudo-Dirac 
         neutrino mass $M_{\zeta_1}$ where $\mu_S$ matrix has been determined for NH light neutrino 
         masses. The resonance peak is at ${M_\zeta}_1\simeq 190$~MeV as shown by the solid line. 
         For comparison, the standard contribution with NH masses is shown by the dashed 
         horizontal line.} 
\label{fig:resonance}
\end{figure}

The predicted nonstandard contributions to $0\nu \beta \beta$ transition are shown in 
the {\it left-panel} for NH and in the {it right-panel} for IH case of Fig. \ref{plot:sm-pseudo-c}.
In view of the ${M_\zeta}_1$-dependent enhancements of $0\nu\beta\beta$ decay rates discussed above
it is tempting to search for the possibility of the lightest pseudo-Dirac neutrino mass which we perform
by the replacement ${M^2_\zeta}_1\rightarrow{M^2_\zeta}_1+i{M_\zeta}_1\Gamma_1$, where $\Gamma_1$ corresponds 
to plausible value of width of the particle. Using, for example, $\Gamma_1\simeq 0.1$~keV, our predictions are
presented by solid curve in Fig.~\ref{fig:resonance} for NH light neutrino masses where the resonant behavior
is clearly exhibited around ${M_\zeta}_1=190$~MeV.

{\bf \small \hspace{-0.5cm} V. DISCUSSIONS AND CONCLUSION:}
In this letter we have shown that in the inverse seesaw framework of the standard model, 
the lightest of the pseudo-Dirac neutrino could be of ${\cal O}(100)$~MeV in concordance 
with tiny left handed neutrino masses and the oscillation data. This pseudo-Dirac neutrino 
mass being in the vicinity of the neutrino virtuality momentum $|p|\simeq 190$~MeV, gives 
very significant non-standard contributions to $0\nu\beta\beta$ decay rates, even far exceeding 
the standard contributions. The Dirac neutrino mass possibly originating from high scale Pati-Salam 
symmetry or SO(10) grand unification, plays a crucial role in determining dominant contributions to 
$0\nu\beta\beta$ decay rates simultaneously with LFV decays with predicted branching ratios 
accessible to on going search experiments. The underlying mechanism provides three distinct platforms 
for its falsifiability {\bf (i)} $0\nu\beta\beta$ decay rates, {\bf (ii)} determination of light pseudo-Dirac 
neutrino mass ${M_\zeta}_1\simeq 120-500$~MeV, and {\bf (iii)} the three predicted branching ratios of 
eqn. (\ref{eqn:branch_pred}). As all our results are applicable in the case of inverse seesaw extended 
supersymmetric standard model, they are also consistent with gauge coupling unification at the 
MSSM-GUT scale, $M_U \simeq 2 \times 10^{16}$ GeV.
The Pati-Salam or SO(10) completion of the model discussed in Sec. IV.B 
will be reported elsewhere in future publication \cite{future-publn}.

{\bf \small \hspace{-0.45cm} ACKNOWLEDGEMENT:}
Ram Lal Awasthi acknowledges the hospitality at Center of Excellence in Theoretical and Mathematical 
Sciences, SOA University where the present work has been completed.

\end{document}